Article In Space without Suit after Joseph v3 6 17 08

# In Outer Space without a Space Suit?*

**Alexander Bolonkin**
1310 Avenue R, #6-F, Brooklyn, NY 11229
T/F 718-339-4563, aBolonkin@juno.com, http://Bolonkin.narod.ru

**Abstract**

The author proposes and investigates his old idea – a living human in space without the encumbrance of a complex space suit. Only in this condition can biological humanity seriously attempt to colonize space because all planets of Solar system (except the Earth) do not have suitable atmospheres. Aside from the issue of temperature, a suitable partial pressure of oxygen is lacking. In this case the main problem is how to satiate human blood with oxygen and delete carbonic acid gas (carbon dioxide). The proposed system would enable a person to function in outer space without a space suit and, for a long time, without food. That is useful also in the Earth for sustaining working men in an otherwise deadly atmosphere laden with lethal particulates (in case of nuclear, chemical or biological war), in underground confined spaces without fresh air, under water or a top high mountains above a height that can sustain respiration.

* Presented to http://arxiv.org on 23 June, 2008.

## Introduction

**Short history.** A fictional treatment of Man in space without spacesuit protection was famously treated by Arthur C. Clarke in at least two of his works, "Earthlight" and the more famous "2001: A Space Odyssey". In the scientific literature, the idea of sojourning in space without complex space suits was considered seriously about 1970 and an initial research was published in [1] p.335 – 336. Here is more detail research this possibility.

**Humans and vacuum.** Vacuum is primarily an asphyxiant. Humans exposed to vacuum will lose consciousness after a few seconds and die within minutes, but the symptoms are not nearly as graphic as commonly shown in pop culture. Robert Boyle was the first to show that vacuum is lethal to small animals. Blood and other body fluids do boil (the medical term for this condition is ebullism), and the vapour pressure may bloat the body to twice its normal size and slow circulation, but tissues are elastic and porous enough to prevent rupture. Ebullism is slowed by the pressure containment of blood vessels, so some blood remains liquid. Swelling and ebullism can be reduced by containment in a flight suit. Shuttle astronauts wear a fitted elastic garment called the Crew Altitude Protection Suit (CAPS) which prevents ebullism at pressures as low as 15 Torr (2 kPa). However, even if ebullism is prevented, simple evaporation of blood can cause decompression sickness and gas embolisms. Rapid evaporative cooling of the skin will create frost, particularly in the mouth, but this is not a significant hazard.

Animal experiments show that rapid and complete recovery is the norm for exposures of fewer than 90 seconds, while longer full-body exposures are fatal and resuscitation has never been successful.[4] There is only a limited amount of data available from human accidents, but it is consistent with animal data. Limbs may be exposed for much longer if breathing is not impaired. Rapid decompression can be much more dangerous than vacuum exposure itself. If the victim holds his breath during decompression, the delicate internal structures of the lungs can be ruptured, causing death. Eardrums may be ruptured by rapid decompression, soft tissues may bruise and seep blood, and the stress of shock will accelerate oxygen consumption leading to asphyxiation.

In 1942, the Nazi regime tortured Dachau concentration camp prisoners by exposing them to vacuum. This was an experiment for the benefit of the German Air Force (Luftwaffe), to determine the human body's capacity to survive high altitude conditions.

Some extremophile microrganisms, such as Tardigrades, can survive vacuum for a period of years.

**Respiration (physiology).** In animal physiology, respiration is the transport of oxygen from the clean air to the tissue cells and the transport of carbon dioxide in the opposite direction. This is in contrast to the biochemical definition of respiration, which refers to cellular respiration: the metabolic process by which an organism obtains energy by reacting oxygen with glucose to give water, carbon dioxide and ATP (energy). Although physiologic respiration is necessary to sustain cellular respiration and thus life in animals, the processes are distinct: cellular respiration takes place in individual cells of the animal, while physiologic respiration concerns the bulk flow and transport of metabolites between the organism and external environment.

In unicellular organisms, simple diffusion is sufficient for gas exchange: every cell is constantly bathed in the external environment, with only a short distance for gases to flow across. In contrast, complex multicellular organisms such as humans have a much greater distance between the environment and their innermost cells, thus, a respiratory system is needed for effective gas exchange. The respiratory system works in concert with a circulatory system to carry gases to and from the tissues.

In air-breathing vertebrates such as humans, respiration of oxygen includes four stages:

- *Ventilation* from the ambient air into the alveoli of the lung.
- *Pulmonary gas exchange* from the alveoli into the pulmonary capillaries.
- *Gas transport* from the pulmonary capillaries through the circulation to the peripheral capillaries in the organs.
- *Peripheral gas exchange* from the tissue capillaries into the cells and mitochondria.

Note that ventilation and gas transport require energy to power mechanical pumps (the diaphragm and heart respectively), in contrast to the passive diffusion taking place in the gas exchange steps.

Respiratory physiology is the branch of human physiology concerned with respiration.

**Respiration system.** In humans and other mammals, the respiratory system consists of the airways, the lungs, and the respiratory muscles that mediate the movement of air into and out of the body. Within the alveolar system of the lungs, molecules of oxygen and carbon dioxide are passively exchanged, by diffusion, between the gaseous environment and the blood. Thus, the respiratory system facilitates oxygenation of the blood with a concomitant removal of carbon dioxide and other gaseous metabolic wastes from the circulation. The system also helps to maintain the acid-base balance of the body through the efficient removal of carbon dioxide from the blood.

**Circulation.** The right side of the heart pumps blood from the right ventricle through the pulmonary semilunar valve into the pulmonary trunk. The trunk branches into right and left pulmonary arteries to the pulmonary blood vessels. The vessels generally accompany the airways and also undergo numerous branchings. Once the gas exchange process is complete in the pulmonary capillaries, blood is returned to the left side of the heart through four pulmonary veins, two from each side. The pulmonary circulation has a very low resistance, due to the short distance within the lungs, compared to the systemic circulation, and for this reason, all the pressures within the pulmonary blood vessels are normally low as compared to the pressure of the systemic circulation loop.

Virtually all the body's blood travels through the lungs every minute. The lungs add and remove many chemical messengers from the blood as it flows through pulmonary capillary bed. The fine capillaries also trap blood clots that have formed in systemic veins.

**Gas exchange**. The major function of the respiratory system is gas exchange. As gas exchange occurs, the acid-base balance of the body is maintained as part of homeostasis. If proper ventilation is not maintained, two opposing conditions could occur: 1) respiratory acidosis, a life threatening condition, and 2) respiratory alkalosis.

Upon inhalation, gas exchange occurs at the alveoli, the tiny sacs which are the basic functional component of the lungs. The alveolar walls are extremely thin (approx. 0.2 micrometres), and are permeable to gases. The alveoli are lined with pulmonary capillaries, the walls of which are also thin enough to permit gas exchange.

**Membrane oxygenator.** A membrane oxygenator is a device used to add oxygen to, and remove carbon dioxide from the blood. It can be used in two principal modes: to imitate the function of the lungs in cardiopulmonary bypass (CPB), and to oxygenate blood in longer term life support, termed Extracorporeal membrane oxygenation, ECMO. A membrane oxygenator consists of a thin gas permeable membrane separating the blood and gas flows in the CPB circuit; oxygen diffuses from the gas side into the blood, and carbon dioxide diffuses from the blood into the gas for disposal.

The introduction of microporous hollow fibres with very low resistance to mass transfer revolutionised design of membrane modules, as the limiting factor to oxygenator performance became the blood resistance [Gaylor, 1988]. Current designs of oxygenator typically use an extraluminal flow regime, where the blood flows outside the gas filled hollow fibres, for short term life support, while only the <u>homogeneous</u> membranes are approved for long term use.

**Heart-lung machine.** The heart-lung machine is a mechanical pump that maintains a patient's blood circulation and oxygenation during heart surgery by diverting blood from the venous system, directing it through tubing into an artificial lung (oxygenator), and returning it to the body. The oxygenator removes carbon dioxide and adds oxygen to the blood that is pumped into the arterial system.

**Space suit**. A space suit is a complex system of garments, equipment and environmental systems designed to keep a person alive and comfortable in the harsh environment of outer space. This applies to extra-vehicular activity (EVA) outside spacecraft orbiting Earth and has applied to walking, and riding the Lunar Rover, on the Moon.

Some of these requirements also apply to pressure suits worn for other specialized tasks, such as high-altitude reconnaissance flight. Above Armstrong's Line (~63,000 ft/~19,000 m), pressurized suits are needed in the sparse atmosphere. Hazmat suits that superficially resemble space suits are sometimes used when dealing with biological hazards.

A conventional space suit must perform several functions to allow its occupant to work safely and comfortably. It must provide: A stable internal pressure, Mobility, Breathable oxygen, Temperature regulation, Means to recharge and discharge gases and liquids , Means of collecting and containing solid and liquid waste, Means to maneuver, dock, release, and/or tether onto spacecraft.

**Operating pressure.** Generally, to supply enough oxygen for respiration, a spacesuit using pure oxygen must have a pressure of about 4.7 psi (32.4 kPa), equal to the 3 psi (20.7 kPa) partial pressure of oxygen in the Earth's atmosphere at sea level, plus 40 torr (5.3 kPa) $CO_2$ and 47 torr (6.3 kPa) water vapor pressure, both of which must be subtracted from the alveolar pressure to get alveolar oxygen partial pressure in 100% oxygen atmospheres, by the alveolar gas equation. The latter two figures add to 87 torr (11.6 kPa, 1.7 psi), which is why many modern spacesuits do not use 3 psi, but 4.7 psi (this is a slight overcorrection, as alveolar partial pressures at sea level are not a full 3 psi, but a bit less). In spacesuits that use 3 psi, the astronaut gets only 3 - 1.7 = 1.3 psi (9 kPa) of oxygen, which is about the alveolar oxygen partial pressure attained at an altitude of 6100 <u>ft</u> (1860 m) above sea level. This is about 78% of normal sea level pressure, about the same as pressure in a commercial passenger jet aircraft, and is the realistic lower limit for safe ordinary space suit pressurization which allows reasonable work capacity.

Movements are seriously restricted in the suits, with a mass of more than 110 kilograms each (Shenzhou 7 space suit). The current space suits are very expensive. Flight-rated NASA spacesuits cost about $22,000,000. While other models may be cheaper, sale is not currently open even to the wealthy public. Even if spaceflight were free (a huge if) a person of average means could not afford to walk in space or upon other planets.

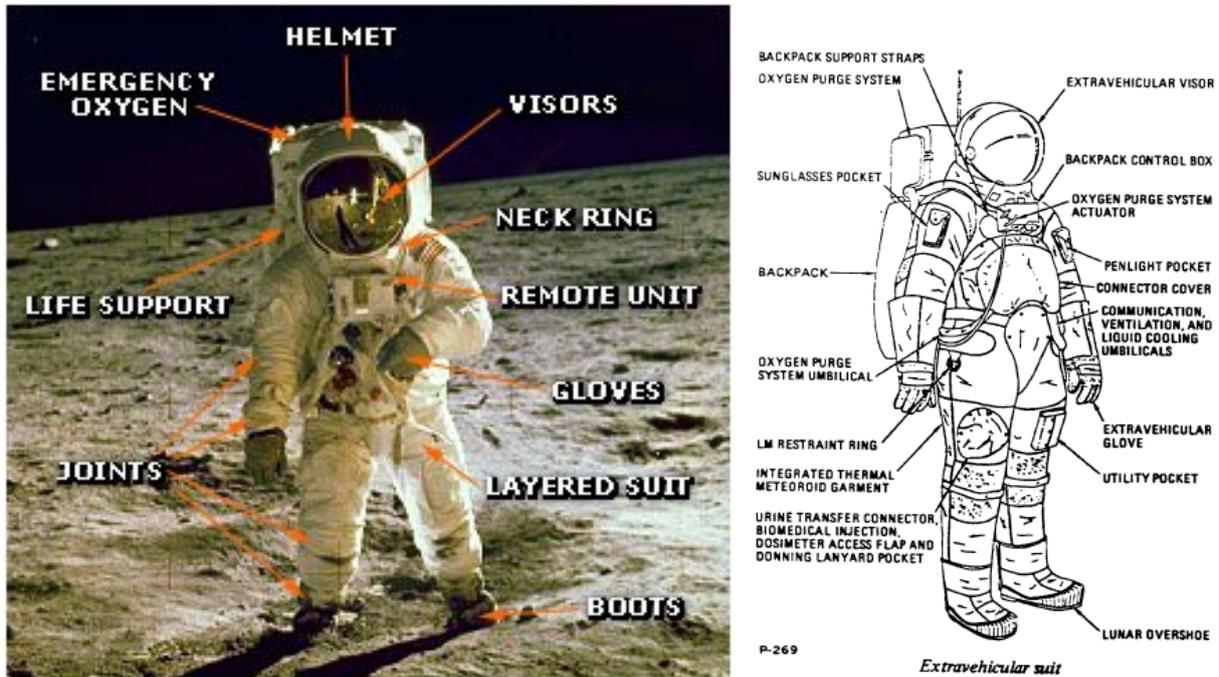

**Fig. 1**. a. Apollo 11 A7L space suit. b. Diagram showing component parts of A7L space suit.

## Brief Description of Innovation

A space suit is a very complex and expensive device (Fig. 1). Its function is to support the person's life, but it makes an astronaut immobile and slow, prevents him or her working, creates discomfort, does not allows eating in space, have a toilet, etc. Astronauts need a space ship or special space habitat located not far from away where they can undress for eating, toilet activities, and rest.

Why do we need a special space suit in outer space? There is only one reason – we need an oxygen atmosphere for breathing, respiration. Human evolution created lungs that aerates the blood with oxygen and remove carbon dioxide. However we can also do that using artificial apparatus. For example, doctors, performing surgery on someone's heart or lungs connect the patient to a heart – lung machine that acts in place of the patent's lungs or heart.

We can design a small device that will aerate the blood with oxygen and remove the carbon dioxide. If a tube from the main lung arteries could be connected to this device, we could turn on (off) the artificial breathing at any time and enable the person to breathe in a vacuum (on an asteroid or planet without atmosphere) in a degraded or poisonous atmosphere, or under water, for a long time. In space we can use a conventional Earth manufacture oversuit (reminiscent of those used by workers in semiconductor fabs) to protect us against solar ultraviolet light.

The sketch of device which saturates the blood with oxygen and removes the carbon dioxide is presented in fig.2. The Heart-Lung machines are widely used in current surgery.

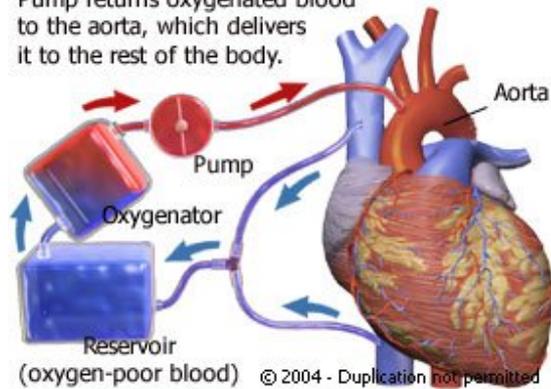

**Fig.2.** Principal sketch of heart-Lung Machine

The main part of this device is oxygenator, which aerates the blood with oxygen and removes the carbon dioxide. The principal sketch of typical oxygenator is presented in fig. 3.

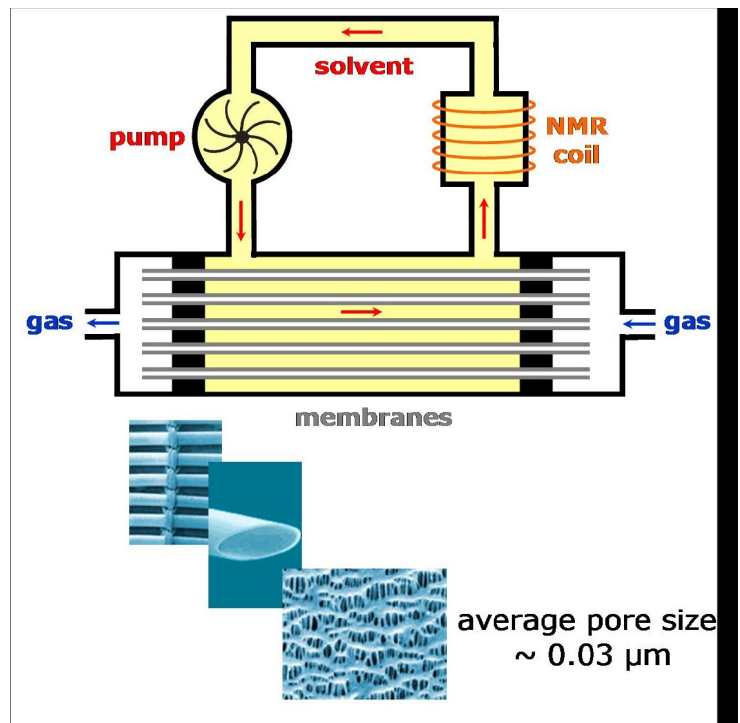

**Fig.3.** Principal sketch of oxygenator.

Current oxygenator is shown in Fig. 4.

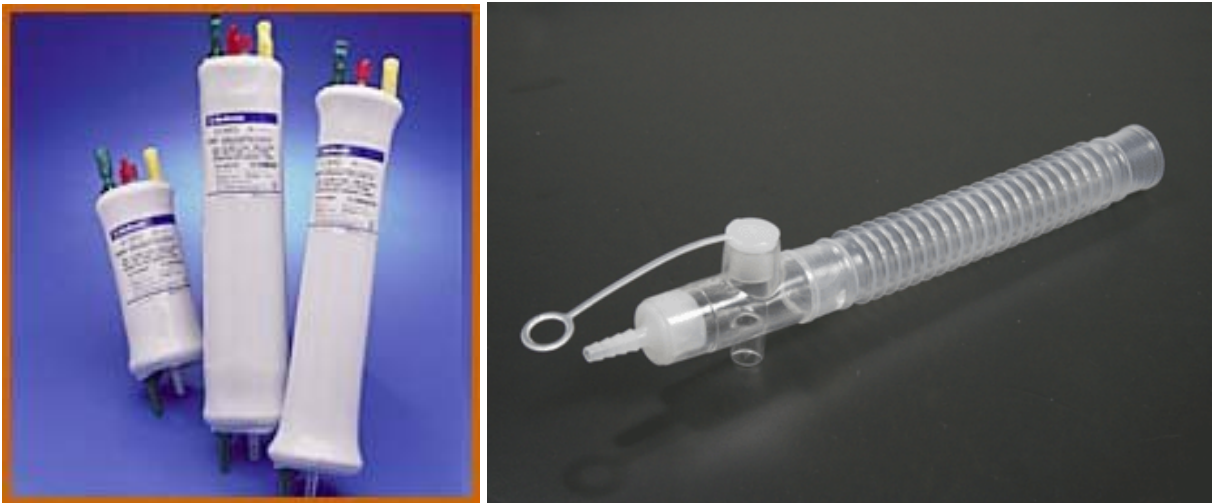

**Fig. 4**. Oxygenators

The human blood circulatory system is shown in Fig. 5.

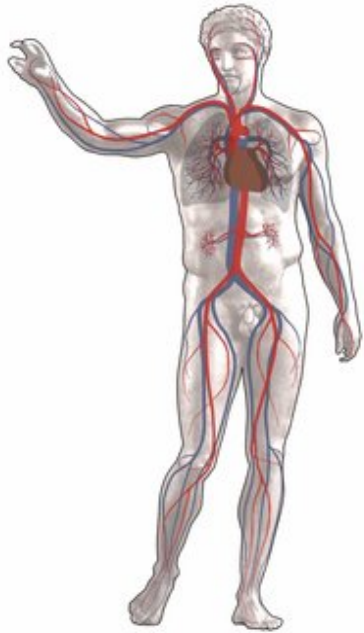

**Fig.5.** The human circulatory system. Red indicates oxygenated blood, blue indicates deoxygenated.

The **circulatory system** is an organ system that moves nutrients, gases, and wastes to and from cells, helps fight diseases and helps stabilize body temperature and pH to maintain homeostasis. This system may be seen strictly as a blood distribution network, but some consider the circulatory system as composed of the cardiovascular system, which distributes blood, and the lymphatic system, which distributes lymph. While humans, as well as other vertebrates, have a closed cardiovascular system (meaning that the blood never leaves the network of arteries, veins and capillaries), some invertebrate groups have an open cardiovascular system. The most primitive animal phyla lack circulatory systems. The lymphatic system, on the other hand, is an open system.

The main components of the human circulatory system are the heart, the blood, and the blood vessels. The circulatory system includes: the pulmonary circulation, a "loop" through the lungs where blood is oxygenated; and the systemic circulation, a "loop" through the rest of the body to provide oxygenated blood. An average adult contains five to six quarts (roughly 4.7 to 5.7 liters) of blood, which consists of plasma that contains red blood cells, white blood cells, and platelets.

Two types of fluids move through the circulatory system: blood and lymph. The blood, heart, and blood vessels form the cardiovascular system. The lymph, lymph nodes, and lymph vessels form the lymphatic system. The cardiovascular system and the lymphatic system collectively make up the circulatory system.

The simplest form of intravenous access is a <u>syringe</u> with an attached **hollow needle.** The needle is inserted through the skin into a vein, and the contents of the syringe are injected through the needle into the bloodstream. This is most easily done with an arm vein, especially one of the metacarpal veins. Usually it is necessary to use a constricting band first to make the vein bulge; once the needle is in place, it is common to draw back slightly on the syringe to aspirate blood, thus verifying that the needle is really in a vein; then the constricting band is removed before injecting.

When man does not use the outer air pressure in conventional space suite, he not has opposed internal pressure except the heart small pressure in blood. The skin vapor easy stop by film clothes or make small by conventional clothes.

The current lung devices must be re-designed for space application. These must be small, light, cheap, easy in application (using hollow needles, no operation (surgery)!), work a long time in field conditions. Wide-ranging space colonization by biological humanity is impossible without them.

### Artificial Nutrition.

Application of offered devices gives humanity a unique possibility to be a long time without conventional nutrition. Many will ask, "who would want to live like that?" But in fact many crew members, military, and other pressured personnel routinely cut short what most would consider normal dining routines. And there are those morbidly obese people for whom dieting is difficult exactly because (in an unfortunate phrase!) many can give up smoking 'cold turkey', but few can give up 'eating cold turkey'! Properly 'fed' intravenously, a person could lose any amount of excess weight he needed to, while not suffering hunger pains or the problems the conventional eating cycle causes. It is known that people in a coma may exist some years in artificial nutrition inserted into blood. Let us consider the current state of the art.

**Total parenteral nutrition** (TPN), is the practice of feeding a person intravenously, bypassing the usual process of eating and digestion. The person receives nutritional formulas containing <u>salts</u>, glucose, amino acids, lipids and added vitamins.

Total parenteral nutrition (TPN), also referred to as Parenteral nutrition (PN), is provided when the gastrointestinal tract is nonfunctional because of an interruption in its continuity or because its absorptive capacity is impaired. It has been used for comatose patients, although enteral feeding is usually preferable, and less prone to complications. Short-term TPN may be used if a person's digestive system has shut down (for instance by Peritonitis), and they are at a low enough weight to cause concerns about nutrition during an extended hospital stay. Long-term TPN is occasionally used to treat people suffering the extended consequences of an accident or surgery. Most controversially, TPN has extended the life of a small number of children born with nonexistent or severely deformed guts. The oldest were eight years old in 2003.

The preferred method of delivering TPN is with a medical infusion pump. A sterile bag of nutrient solution, between 500 mL and 4 L is provided. The pump infuses a small amount (0.1 to 10 mL/hr) continuously in order to keep the vein open. Feeding schedules vary, but one common regimen ramps up the nutrition over a few hours, levels off the rate for a few hours, and then ramps it down over a few more hours, in order to simulate a normal set of meal times.

Chronic TPN is performed through a central intravenous catheter, usually in the subclavian or jugular vein. Another common practice is to use a PICC line, which originates in the arm, and extends to one of the central veins, such as the subclavian. In infants, sometimes the umbilical vein is used.

Battery-powered ambulatory infusion pumps can be used with chronic TPN patients. Usually the pump and a small (100 ml) bag of nutrient (to keep the vein open) are carried in a small bag around the waist or on the shoulder. Outpatient TPN practices are still being refined.

Aside from their dependence on a pump, chronic TPN patients live quite normal lives.

Central IV lines flow through a catheter with its tip within a large vein, usually the superior vena cava or inferior vena cava, or within the right atrium of the heart.

There are several types of catheters that take a more direct route into central veins. These are collectively called *central venous lines*.

In the simplest type of central venous access, a catheter is inserted into a subclavian, internal jugular, or (less commonly) a femoral vein and advanced toward the heart until it reaches the superior vena cava or right atrium. Because all of these veins are larger than peripheral veins, central lines can deliver a higher volume of fluid and can have multiple lumens.

Another type of central line, called a Hickman line or Broviac catheter, is inserted into the target vein and then "tunneled" under the skin to emerge a short distance away. This reduces the risk of infection, since bacteria from the skin surface are not able to travel directly into the vein; these catheters are also made of materials that resist infection and clotting.

## Testing

The offered idea may be easily investigated in animals on Earth by using currently available devices. The experiment includes the following stages:
1) Using a hollow needle, the main blood system of a good healthy animal connects to a current heart-lung machine.
2) The animal is inserted under a transparent dome and air is gradually changed to a neutral gas (for example, nitrogen). If all signs are OK we may proceed to the following stage some days later.
3) The animal is inserted under a transparent dome and air is slowly (tens of minutes) pumped out. If all signs are OK we may start the following stage.
4) Investigate how long time the animal can be in vacuum? How quick we can decompress and compress? How long the animal may live on artificial nutrition? And so on.
5) Design the lung (oxygenator) devices for people which will be small, light, cheap, reliable, safe, which delete gases from blood (especially those that will cause 'bends' in the case of rapid decompression, and work on decreasing the decompressing time).
6) Testing the new devices on animals then human volunteers.

## Advantages of offered system.

The offered method has large advantages in comparison with space suits:
1) The lung (oxygenator) devices are small, light, cheaper by tens to hundreds times than the current space suit.
2) It does not limit the activity of a working man.
3) The working time increases by some times. (less heat buildup, more supplies per a given carry weight, etc)
4) It may be widely used in the Earth for existing in poison atmospheres (industry, war), fire, rescue operation, under water, etc.
5) Method allows permanently testing (controlling) the blood and immediately to clean it from any poison and gases, wastes, and so on. That may save human lives in critical medical situations and in fact it may become standard emergency equipment.
6) For quick save the human life.
7) Pilots for high altitude flights.

8) The offered system is a perfect rescue system because you turn off from environment and exist INDEPENDENTLY from the environment. (Obviously excluding outside thermal effects, fires etc—but for example, many fire deaths are really smoke inhalation deaths; the bodies are often not burned to any extent. In any case it is much easier to shield searing air from the lungs if you are not breathing it in!)

**Conclusion.**

The author proposes and investigates his old idea – a living human in space without the encumbrance of a complex space suit. Only in this condition can biological humanity seriously attempt to colonize space because all planets of Solar system (except the Earth) do not have suitable atmospheres. Aside from the issue of temperature, a suitable partial pressure of oxygen is lacking. In this case the main problem is how to satiate human blood with oxygen and delete carbonic acid gas (carbon dioxide). The proposed system would enable a person to function in outer space without a space suit and, for a long time, without food. That is useful also in the Earth for sustaining working men in an otherwise deadly atmosphere laden with lethal particulates (in case of nuclear, chemical or biological war), in underground confined spaces without fresh air, under water or a top high mountains above a height that can sustain respiration. There also could be numerous productive medical uses.

*Acknowledgement*

The author wishes to acknowledge Joseph Friedlander for correcting the English and offering useful advice.